\documentclass[aps,prl,preprint,amsmath,amssymb]{revtex4}

\usepackage{graphicx}
\begin{document}

\newcommand{\rf}[1]{(\ref{#1})}

\newcommand{\bfomega}{ \mbox{\boldmath{$\omega$}}}

\title{Experimental measurement of an effective temperature 
for jammed granular materials}

\author{Chaoming Song} 
\author{Ping Wang} 
\author{Hern\'an A. Makse}

\affiliation { Levich Institute and Physics Department
\\ City College
of New York
\\ New York, NY 10031, US}

\date{ Proc. Nat. Acad. Sci. 102, 2299-2304 (2005)}

\begin{abstract}

A densely packed granular system 
is an example of an out-of-equilibrium system in
the jammed state.  It has been a longstanding problem to determine
whether this class of systems can be described by concepts arising
from equilibrium statistical mechanics, such as an ``effective
temperature'' and ``compactivity''. 
The measurement of the effective temperature is
realized in the laboratory by slowly shearing a closely-packed
ensemble of spherical beads confined by an external pressure in a
Couette geometry. All the probe particles considered in this study,
 independent of their
characteristic features, equilibrate at the same temperature, given by
the packing density of the system.  

\end{abstract}
\maketitle

\clearpage

The application of the 
concept of ``effective temperature'' to out-of-equilibrium
systems, which 
allows the extension of ideas from equilibrium statistical mechanics
such as the fluctuation-dissipation relation \cite{landau}
to nonequilibrium systems,
has been extensively debated in the literature \cite{ckp,liu1,mehta,coniglio,wolf}.
Previous experimental
measurements have been performed on structural glasses \cite{grigera},
colloidal suspensions \cite{bellon}, spin glasses \cite{herisson},
highly agitated granular matter \cite{danna} and a fluidized particle
\cite{durian}.

Until present, the only evidence of its significance in describing
closely packed (jammed) granular media \cite{liu1} 
has emerged from
computer simulations of granular materials and other 
soft-matter systems
and from 
analogies with glassy
systems \cite{kurchan,liu3,bklm,sciortino,barrat,mehta3,ono,coniglio2,mk}. 
 The existence of jammed reversible states
in granular materials 
has been also suggested
by compaction experiments employing tapping, oscillatory compression
or sound propagation as the external perturbation
\cite{nowak,bideau,cavendish,edwards2,bideau2}.  
However, macroscopic
variables,
 such as the effective temperature for highly packed granular matter, 
have not been previously
measured in the laboratory.
This line of research has led
to the design of the experiment which we are about to describe.


In this paper we present experimental evidence 
suggesting the existence of a well-defined
effective temperature for slowly sheared granular materials
very close to jamming.
The particle
trajectories in the sheared system 
yield the diffusivity and the mobility from which the
temperature is deduced using a fluctuation-dissipation
relation.  All the particles considered in this study 
equilibrate at the same temperature, 
which is in turn
independent of the slow shear rate, thus suggesting the condition of a
physical variable to describe the jammed system.

\begin{figure}
\centering 
\resizebox{16cm}{!}{\includegraphics{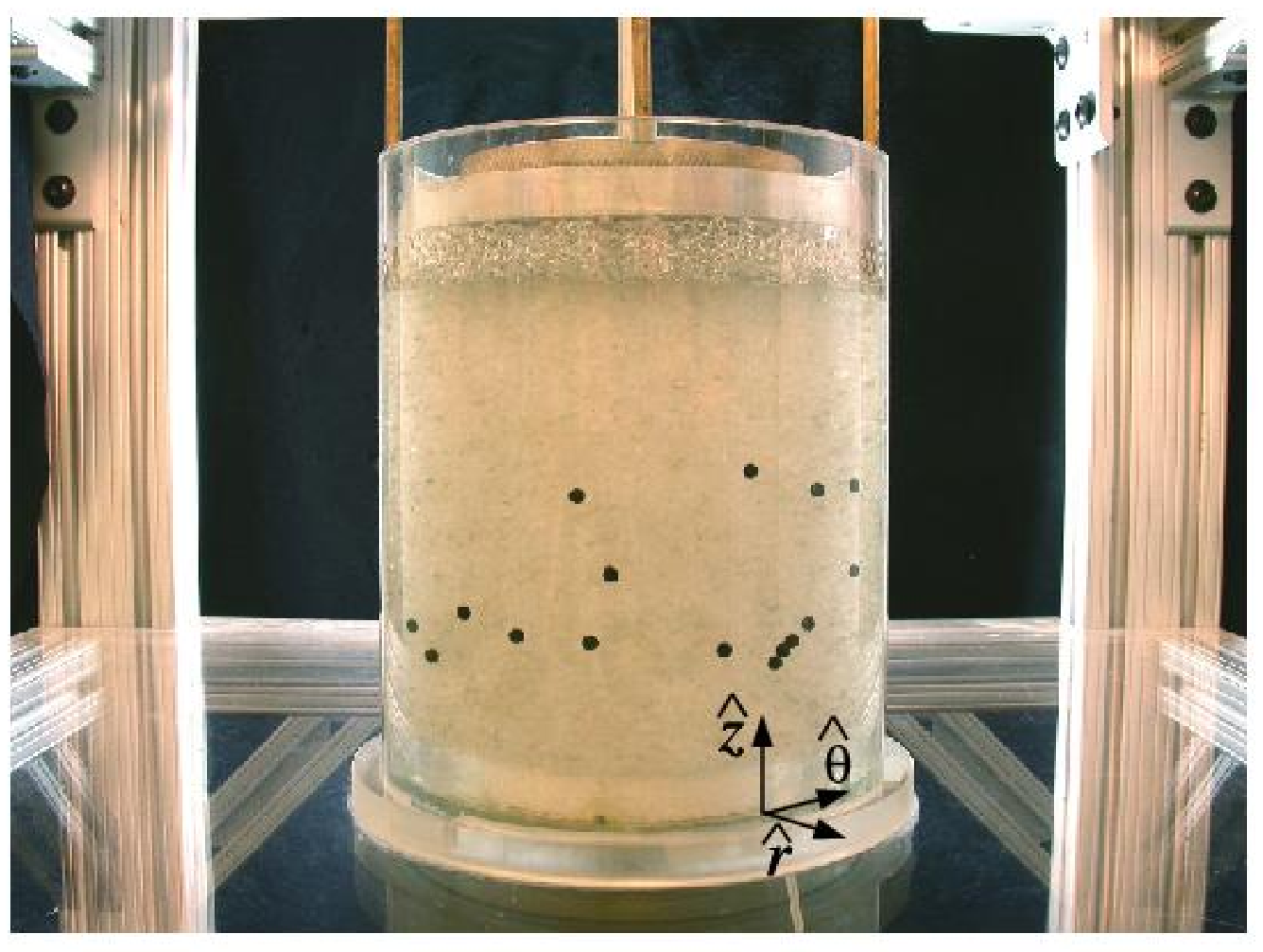}}
  \caption{}
\label{couette}
\end{figure}

{\bf Experiments.--} The experimental test 
involves using
observational techniques to monitor the evolution of the particulate
packing as it explores the available
configurations.
 The different packing configurations are investigated using
quasi-static shear in a Couette cell geometry, depicted in Fig.
\ref{couette}.


%

The particulate system is confined between an inner and an outer
cylinder. The inner cylinder of the cell is slowly rotated by a motor
while the outer cylinder is fixed and transparent for
visualization. The walls of both cylinders are roughened by gluing a
layer of
particles to provide shear motion to the assembly, avoiding
wall-slip. The grains are compactified by the application of an
external pressure of a specific value (typically 386 Pa), introduced
by a moving piston at the top of the granular matrix. We use a narrow
gap Couette cell of the order of 7 particle diameters wide in order
to avoid the formation of bulk shear bands
\cite{nedderman,drake,mueth,veje,utter}.

Since a granular assembly is optically impenetrable by its very
nature the first task consists in creating a transparent sample.
This is achieved by refractive index matching transparent particles
with a suitable suspending solution. The presence of the solution
reduces friction between the particles, nevertheless 
the ensemble remains jammed
throughout the experiment by the  application of the external force via the
piston.  It is important to note that the liquid only partially fills
the cell (see Figure \ref{couette}).
In this way the pressure of the piston is transmitted only to the
granular particles and not to the fluid.

Hydrodynamic effects from the partial filling of the cell are 
avoided by the extremely slow rotational
speeds applied to our system.
However, the fluid may modify
the interaction between the particles, for instance, by reducing
friction. Thus, the transport coefficients
 measured in our experiments will not be the same
as those of a dry system. 
However, 
the key feature of the system
is that it is closely packed, which is hereby satisfied.
The random motion of the particles is due to the `jamming' forces
exerted by the enduring contacts of all the neighboring particles
which renders the problem nontrivial.

The successful packing consists of a 1:1 mixture of two different
sizes of spherical Poly-methyl methacrylate (acrylic) particles, of
density $\rho = 1.19$ and index of refraction $n=1.49$. We use two
different packings, containing either 3.17 mm and 3.97 mm diameter
particles (Packing 1) or particles of diameter 3.97 mm and 4.76 mm
(Packing 2). The size ratio ensures that crystallization is
avoided. The mixture of particles is immersed in a solution of
approximately 74\% weight fraction of cyclohexyl bromide and 26\%
decalin \cite{weeks}, matching not only the refractive index but more
importantly the density of the acrylic particles. The density matching
fluid eliminates pressure gradients associated with gravity in the
vertical direction. This step avoids problems encountered in previous
tests of compactivity \cite{nowak} and other effects such as
convection and size segregation such as the Brazil nut effect inside
the cell \cite{behringer}.

We follow the trajectories of tracer particles in the sample bulk
to obtain 
the diffusion of the tracers and the
response function (mobility) to an external force within the
structure. These measurements lead to the 
effective
temperature via a fluctuation-dissipation relation generalized to
granular media. If the resulting effective temperature is 
a physical variable of the system, 
it will be
independent of the
properties of the tracer particles-- a necessary but not sufficient condition.
It is to this end that we
contribute experimental results.
We note that to evaluate the broader thermodynamic meaning of the
temperature
it would be needed to examine 
whether different measures of temperature agree as well.
Such tests could be performed, for instance, by measuring the temperature
for different observables in the system.

The tracer particles must experience a constant force, in response to
which the mobility can be measured. Tracers of a different density to
the acrylic particles are then added to the packing, as shown in 
Fig. \ref{couette}.
Two types of tracers, made of nylon ($\rho = 1.12$) and delrin ($\rho
= 1.36$), are employed.
The role of the tracer in the system is to explore the different
packing configurations, and 
the size of the tracers is chosen, accordingly, 
of similar size as the  background particles.
If the tracers are much larger or smaller than the background particles
new physics would be brought into the problem.
For instance if a tracer is small
enough to fall into the voids of the other particles
then ``percolation effects'' \cite{drahun} 
would prevail and the displacements
would be larger than those predicted by the effective
temperature. 
Such a tracer would no longer prove
the background particle network, but would 
instead have different dynamics and 
test other interesting effects which 
cannot be captured by the present formalism.

{\bf Results.---}
The first experiment uses Packing 1 with 20 tracers of  3.97 
mm nylon
beads. The Couette cell is sheared at very slow frequencies 
$f = 2.4$ mHz defining the external shear-rate $\dot \gamma_e = 
2\pi f r_1/(r_2-r_1)= 0.048$
1/s, where $r_1=50.8$ mm and $r_2=66.7$ mm are the radii of the inner
and outer cylinders, respectively.
We require a very slow shear rate so that the system is 
close-packed
at all times. 
The $(r(t),\theta(t),z(t))$ coordinates of the tracers are obtained by
analyzing the images acquired by four digital 
cameras surrounding the shear
cell [($r,\theta$) are obtained
only at the overlaps of the cameras]. 
In the following we first present results for the $z$ direction since
this is the only direction where the temperature can be calculated 
with the present set-up
(the external force acts only vertically).
We discuss the  diffusivities in the 
$\theta$ and $r$ directions at the end of this section.

\begin{figure}
\centerline {\resizebox{14.cm}{!}{\includegraphics{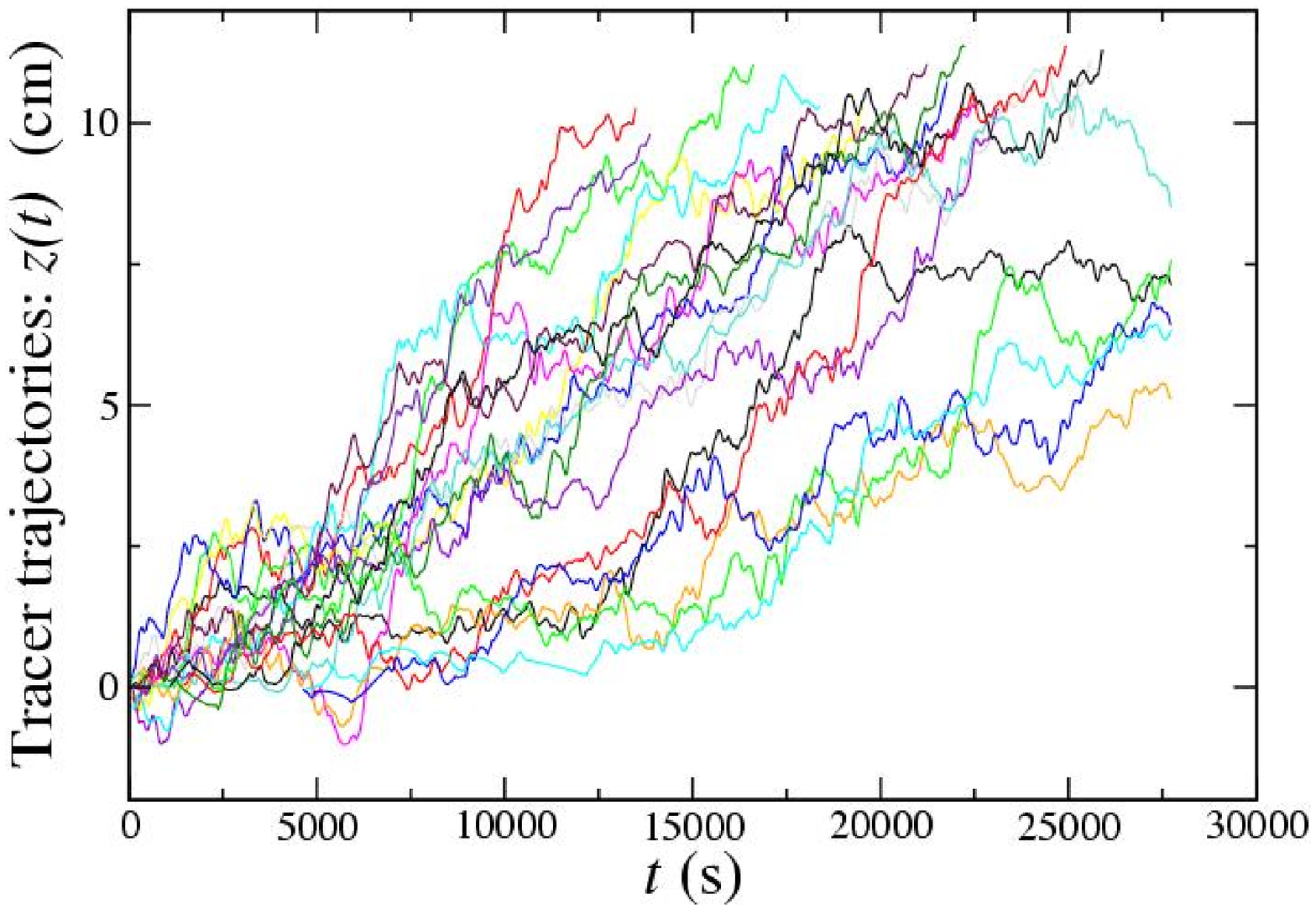}}}
\caption{}
\label{trajectories}
\end{figure}

The resulting vertical trajectories of the tracers $z(t)$ are
depicted in Fig.  \ref{trajectories} showing that the nylon tracers not only
diffuse, but also move with a constant average velocity to the top of
the cell.
We confine the measurements of the tracer fluctuations
away from the inner rotating cylinder
to avoid boundary effects and where
the average tangential velocity of the tracers,
$v_{\theta}(r)$, can be  
approximated linearly 
as $v_{\theta}(r)\approx-\dot\gamma_l r$ with 
a constant local shear rate $\dot\gamma_l = 0.021$ 1/s. This 
ensures that the diffusivity (which depends on the local shear rate)
remains approximately constant in the radial direction.

\begin{figure}
\centering{{\bf a} \hspace{.5cm}
\resizebox{14cm}{!}{\includegraphics{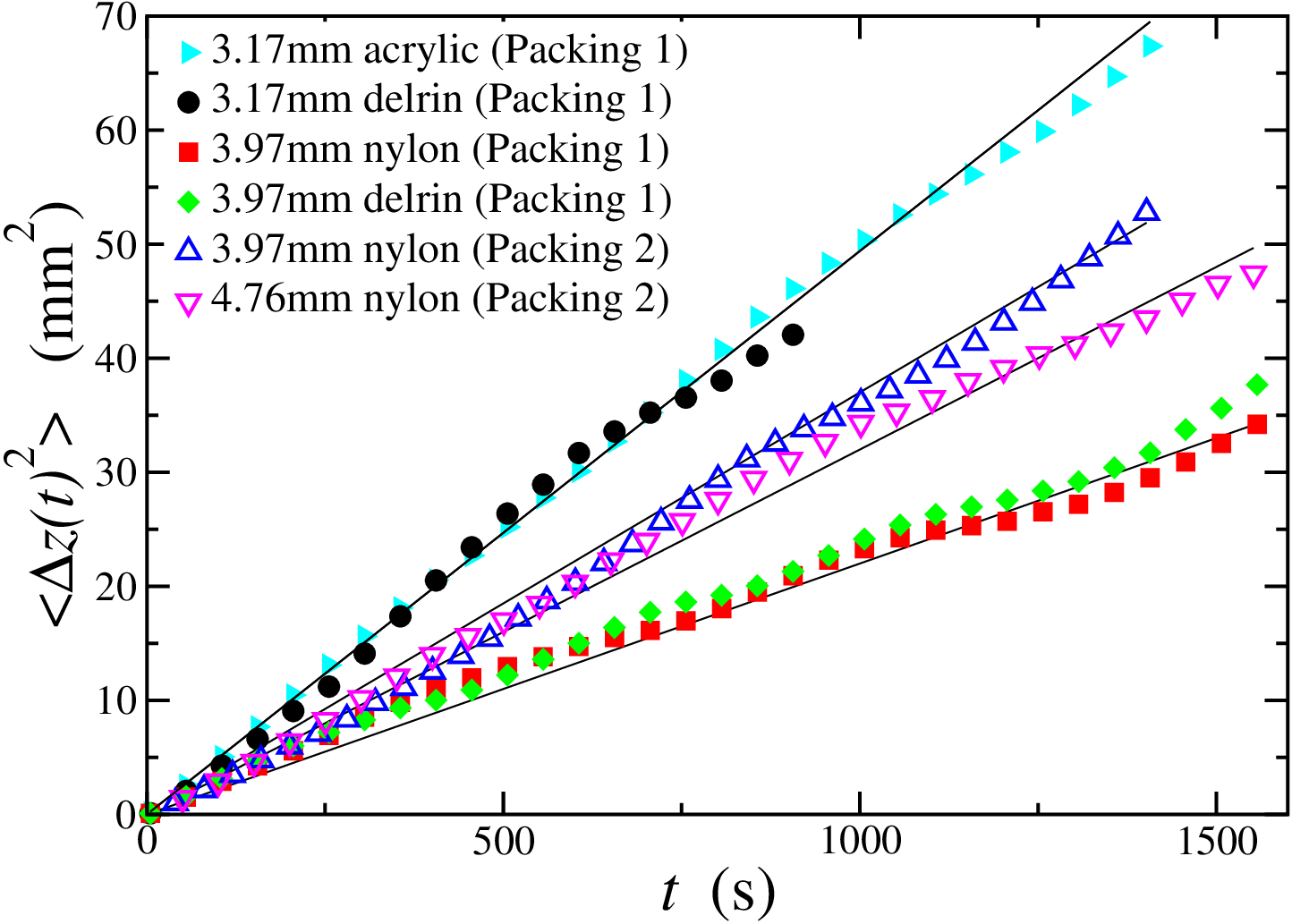}}
}
\vspace{2cm} 

\centering{ {\bf b}  \resizebox{14cm}{!}{\includegraphics{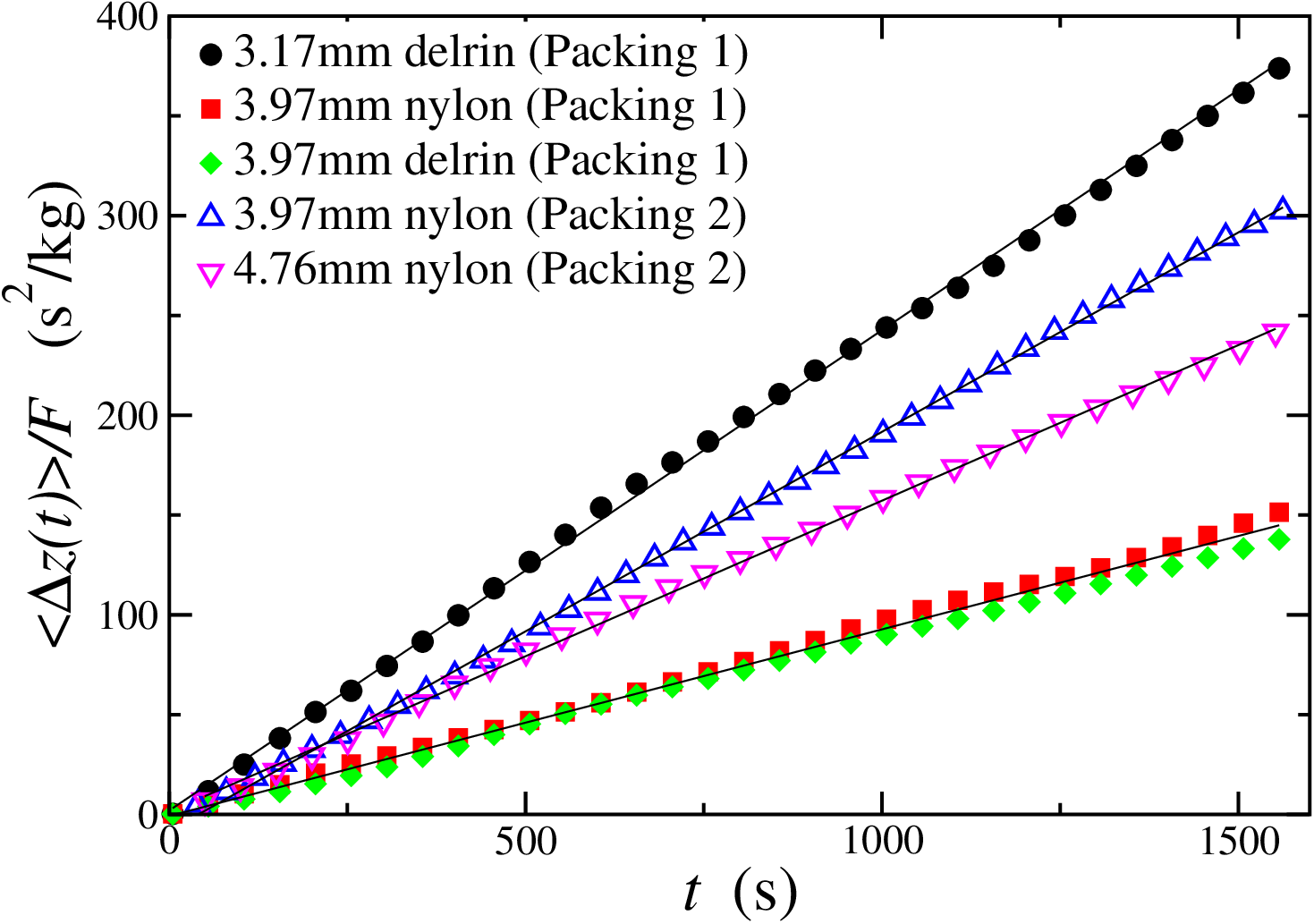}}}
\caption{}
\label{mobility}
\end{figure}

The statistical analysis of the particle displacements $\Delta z(t) =
z(t+t_0) - z(t_0)$ 
%
reveals a Gaussian distribution
which broadens with time, as seen in Fig. \ref{gauss}.  The rms
fluctuations grow linearly for sufficiently long times 
(see Fig. \ref{mobility}a):
\begin{equation}
\langle[z(t+t_0) - z(t_0)]^2\rangle \sim 2 D t,
\end{equation}
where $D$ is the self-diffusion constant. 
For the 3.97 mm tracer we obtain
$D_{\mbox{\scriptsize 3.97mm}} = ( 1.1 \pm 0.1 )\times 10^{-8}$ m$^2$/s.

The average $\langle\cdots\rangle$ 
denotes ensemble average over the tracers and over the initial time $t_0$.
In practice we employ the common method (see Chapter 5.3 in \cite{rapaport})
of performing an average over the time $t_0$ to measure transport coefficients
by splitting the trajectory of a single tracer into a series of trajectories
starting at evenly spaced time intervals $t_0$.
The diffusion constant is then obtained by averaging not only
over the tracers but also over the initial time intervals.
This common technique allows us to 
obtain an estimation
of the diffusion constant using 20 tracers  for this particular
system.  We check that the times series
are taken over intervals for which the correlations between measurements
have decayed almost to zero and that our system 
is time-translational invariant,
i.e. does not display ``aging'' \cite{coniglio}.
Moreover, we check that by doubling the number of tracers we obtain the same
result for $D$ for this particular type of tracer, 
indicating that the average diffusion constant is independent
of the number of tracers used to explore the configurations.

Figure \ref{mobility}b shows the mean value of the position of the
tracers extracted from the peak of the Gaussian distribution as a
function of time, thus yielding the mobility $\chi$ as
\begin{equation}
\langle z(t+t_0) - z(t_0)\rangle \sim \chi F t.
\end{equation}
Here $F = (\rho_{a} - \rho_{t}) V g$ is the gravitational force
applied to the tracers due to their density mismatch while $\rho_{a}$
and $\rho_{t}$ are the densities of the acrylic particles and the
tracers respectively, $V$ is the volume of the tracer particle and $g$
the acceleration of gravity.
The value of the mobility for the nylon 3.97mm tracer is
$\chi_{\mbox{\scriptsize 3.97mm}} = ( 9.7 \pm 0.9)
 \times 10^{-2} $ s/kg.
We check that the mobility of the tracers
is constant in the region of measurements.

We notice that  the rms displacements of the tracer particles 
present a systematic 
downward curvature at long times (see Fig. \ref{mobility}a)
This cut-off is due to a finite size effect because 
the trajectories are finite.
The faster the velocity of the 
tracers (as seen in Fig. \ref{mobility}b) 
the shorter the trajectories and the shorter the 
cut-off time.
Thus this effect is more 
evident in the 3.17 mm delrin tracers in Packing 1
which posse the largest mobility, as seen in Fig. \ref{mobility}b,  and
present the shortest cut-off time in the diffusivity,
 as seen in  Fig. \ref{mobility}a. On the other hand, 
the 3.17mm acrylic tracers
in Packing 1 present a larger cut-off since the external gravitational 
force is not applied to them, and they stay in the window of observation
for much longer times.
We would like to point out that the important issue is that the 
cut-off (which is inevitable in any finite time measurement) 
is observed for distances
larger than a few particles diameters, i.e. for distances larger than
the size of the ``cages''.  This ensures that  we are testing
the structural motion of the grains and not the internal motion inside
the cages (see below).

An important task is to determine whether there exists a linear
response regime in 
the system,
which would imply that the mobility
is independent of the external gravitational force as $F\to 0$.  The external
force is varied by changing the density of the tracers of the same
size.
This is realized experimentally in Packing 1 
by the introduction of
delrin tracers of 3.97 mm diameter, the density of which is higher than
that of nylon. The analysis of the trajectories
reveals that the  mobility is 
the same for both tracers so that it is 
independent of the external force, as
shown in Fig.  \ref{mobility}b. 

The important result that we want to emphasize is that 
there is a well-defined linear response regime for small enough external 
forces where the  
mobility becomes independent of the force.
This regime is achieved for the delrin and nylon tracers.
Moreover, we find that nonlinear effects appear for tracers heavier than
delrin,  i.e. the mobility (which is the velocity normalized by the
external force) depends on the external force for large enough forces.
For instance, we find that for a 3.97mm 
ceramic tracer $(\rho=3.28)$
the mobility is $\chi_{\mbox{\scriptsize ceramic}} = ( 2.2 \pm 0.2)
 \times 10^{-2} $ s/kg and for a brass tracer $(\rho=8.4)$,
$\chi_{\mbox{\scriptsize brass}} = ( 1.7 \pm 0.1)
 \times 10^{-2} $ s/kg, smaller than the mobility of the 
nylon and delrin tracers of the same size.
This  behavior is expected since 
if a linear regime exists in the system, it will be valid
only within certain limits.
We also show that the 
diffusion constants of both types of tracers (delrin and nylon)
are approximately the same (see Fig. \ref{mobility}a) confirming that
the external force on the tracers does not affect the diffusion
constant for the small forces used in this study.

According to a Fluctuation-Dissipation relation, it is the diffusivity
and the mobility of the particles which enable the calculation of an
effective temperature, $T_{\mbox{\scriptsize eff}}$, via an Einstein
relation for sheared granular matter:

\begin{equation}
\langle [z(t+t_0) - z(t_0)]^2\rangle = ~ 2 T_{\mbox{\scriptsize eff}}
~ \frac{\langle z(t+t_0) - z(t_0)\rangle }{F}.
\label{fdt}
\end{equation}

A parametric plot, with $t$ as a parameter, of the fluctuations and
responses is produced to yield the linear relationship shown in
Fig. \ref{teff}, the gradient of which gives $2 T_{\mbox{\scriptsize eff}}$.
We obtain for 
the 3.97 mm tracer 
 $T_{\mbox{\scriptsize eff}} = (1.1 \pm 0.1) \times 10^{-7}$ J.
This value is set by a typical energy scale
in the system  \cite{review}, for instance 
$(\rho_{a} - \rho_{t}) g d$, which is the gravitational 
potential energy to move a tracer particle 
a  distance of the particle
diameter $d$.
The corresponding temperature
which would arise from the conversion of this energy into a
temperature via the Boltzmann constant, $k_B$, is
$T_{\mbox{\scriptsize eff}} 
= 2.7 \times 10^{13} k_B T$ at room temperature. This large value is 
expected \cite{review} (and agrees with
computer simulation estimates \cite{mk}) since granular matter is an
athermal system.

An important evidence in examining the physical  meaning of the
effective temperature can be obtained from the following test:
changing the tracer size should give rise to a different
diffusion and mobility but they should nevertheless
lead to the measurement of the same
effective temperature
if  
the system is at ``equilibrium''.  
We next introduce tracers of 3.17 mm diameter in
Packing 1 and repeat the above calculations.  We find that the 3.17 mm
tracers produce a significantly different diffusion and mobility than
their 3.97 mm counterparts as shown in Fig. \ref{mobility} 
($D_{\mbox{\scriptsize 3.17mm}} = ( 2.5 \pm 0.3) \times 10^{-8}$ m$^2$/s
and $\chi_{\mbox{\scriptsize 3.17mm}} = ( 2.4 \pm 0.3) \times 10^{-1}$ s/kg).
In all
cases $D$ and $\chi$ increase with decreasing size of the tracers.
However, the parametric plot of  diffusivity versus mobility 
demonstrates that their effective temperature are approximately the
 same as seen in
Fig.  \ref{teff} with an average value over all tracers of
 $T_{\mbox{\scriptsize eff}} = (1.1 \pm 0.1) \times 10^{-7}$ J.

We further check that the diffusion is not affected by the 
external force by calculating the diffusivity of the nontracers 
particles by dying acrylic tracers and analyzing their
trajectories.
As shown in Fig. \ref{mobility}a 
the diffusion of the acrylic tracers of size
 3.17 mm (for which no external force is applied) 
is the same as the diffusion of the delrin tracers of the same size 
(for which the gravitational force
is applied).

Next we perform  
another
measurement arising from a repeat of the experiment for a
different packing of spherical particles (Packing 2). 
The use of larger particles of
approximately the same size ratio as in Packing 1 still leads to 
the same volume fraction of
particles.
Since  $T_{\mbox{\scriptsize eff}}$ is a 
measure of how
dense the particulate packing is (i.e. a large  $T_{\mbox{\scriptsize eff}}$
implies a
loose configuration, e.g. random loose packing, while a reduced
$T_{\mbox{\scriptsize eff}}$ implies a more compact structure, e.g. 
random close
packing),
it holds to reason that it should be the same for both of the packings
under investigation. 
Indeed, despite the change in their
respective diffusivities and mobilities as shown in Fig.
\ref{mobility}, the two packings measure approximately the same effective
temperature shown in Fig. \ref{teff}.

An assumption in this study is that of the system being
continuously jammed despite the presence of rearrangements under
shear. We show in the inset of Fig. \ref{teff} that
the effective temperature seems to become approximately
constant, as long as
the particulate motion is slow enough such that enduring contacts
prevail. We find that $D\sim \dot\gamma_e$ and $\chi\sim \dot\gamma_e$,
while $T_{\mbox{\scriptsize eff}}=D/\chi$ remains approximately constant
for sufficiently small  $\dot\gamma_e$. 
It is within this quasi-static range where the
 effective temperature could  be identified
with the exploration of the jammed states.  
The slow shear rate regime observed here could be 
the analogous of the 
shear-rate independent regime observed in the behavior of the shear stress
in slowly sheared granular materials \cite{savage,tardos}.
This solid friction-like behavior 
has been studied in the past \cite{savage,tardos}
and occurs when frictional forces and enduring contacts dominate the dynamics.
This regime has been also observed in recent computer simulations 
of the effective temperature of sheared
granular materials \cite{ono}.
Our results are in accordance with these previous studies.

Given that we are dealing with an athermal system in which the notion
of 'bath' temperature
plays no role 
perhaps
further explanation is required in terms of the actual role of
 $T_{\mbox{\scriptsize eff}}$ in describing granular systems.
The length scale on which the particles diffuse over the long time
scale of the experiment is of the order of a few particle diameters
(see Fig. \ref{trajectories} and \ref{mobility}a)
implying that the exploration of the available 
jammed  configurations
takes place by rearrangements of the particles outside their ``cages''. 
The trajectory of the slow moving system
could be mapped onto the successive jammed states that the system
explores. 
Thus,  we may identify
$T_{\mbox{\scriptsize eff}}$ as the variable governing 
this exploration of the
different jammed configurations.

It has been suggested that, under certain experimental conditions of 
reversibility, a
jammed granular system could be amenable to 
a statistical mechanics formulation \cite{edwards,mehta2}.
The main assumption of this statistical formulation
is that the  different jammed configurations are taken to have
the same statistical weight. Thus, 
observables can be obtained as ``flat'' averages of the jammed 
configurations \cite{kurchan,bklm,coniglio2,mk,luck,mehta3}.
The  validity of this assumption 
has been extensively 
debated in the literature (see for instance \cite{coniglio,cavendish}).
Some simulations and analytical work
suggest that the effective temperature obtained by applying 
the extension of FDT to out-of-equilibrium systems
is indeed analogous to performing a flat average over
the configurational space.
Numerically it has been suggested that the
effective temperature can be identified with the 
compactivity introduced in \cite{edwards}, 
arising from the entropy of the packing \cite{kurchan,bklm,mk}.
Moreover, recent work \cite{edwards2,luck,bideau2}
suggests that
the landscape of a granular system in the jamming limit 
is only flat in a
majoritarian sense; i.e., it is overall flat
despite a very prevalent ruggedness on the microscopic scale.

The experimental test of these 
ideas is very  difficult since the 
entropy of the jammed configurations can not be  easily measured in 
experiments.
For instance, using the present experimental setup it is not possible
to obtain an estimate of the compactivity from entropic
considerations.
Therefore, 
more tests are needed to fully explore the thermodynamic
meaning of the effective temperature and its relation to this 
statistical mechanic framework.




In
contrast to the measurement of the temperature of the slow modes
as exemplified by $T_{\mbox{\scriptsize eff}}$,
we also measure the temperature of the fast modes as
given by the instantaneous rms fluctuations of the 
velocity of the particles. 
This kinetic granular temperature is smaller than 
$T_{\mbox{\scriptsize eff}}$ 
and
differs for each tracer indicating that it is not governed by the same
statistics. Similar results have been obtained in experiments
of vibrated granular gases \cite{menon}.
The significance of this results is  that
there are other modes of relaxation (the fast modes) which are
governed by a different temperature. This result is analogous to what
is found in models of glasses and 
computer simulations of molecular  glasses (see for
instance Refs. \cite{ckp,liu1,mehta,coniglio,wolf}). 
In these models, in the glassy phase
the bath temperature is found to control the fast modes of
relaxation 
and a different 
effective temperature is
found for the slow modes of relaxation. Analogously,
 we find a granular temperature for the fast
modes and an effective temperature for the slow modes of
relaxation.
The fact that different temperatures 
may govern different modes of relaxation in the system
may imply 
that something more
complicated than a Boltzmann approach
may be needed to describe granular
matter.

\begin{figure}
\centerline{
\resizebox{14.cm}{!}{\includegraphics{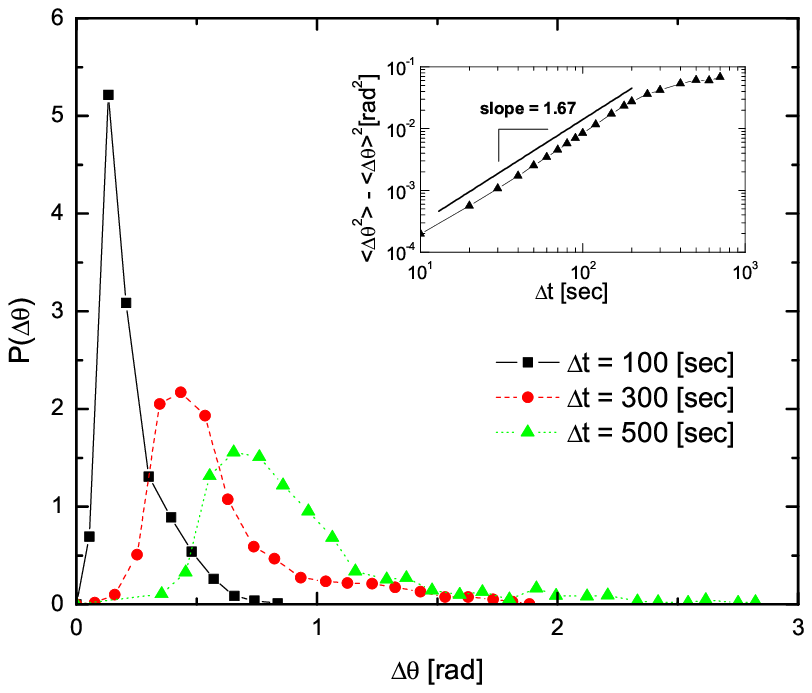}} 
}
\caption{}
\label{sup2}
\end{figure}

Finally, we 
present the analysis of the data in the $\theta$ and in the $r$ directions.
Figure \ref{sup2} shows the results of the  probability distribution of
the displacements  $\Delta \theta(t)$ in the angular  direction.
 The data corresponds to the 3.17 mm delrin tracers in Packing 1 and
is taken every 200 s.  We find that the probability  distributions 
display an asymmetric tail in the direction of the flow.
This extra spreading is
known as the Taylor dispersion \cite{taylor}.

Taylor dispersion appears when diffusion 
couples with the gradient of the flow giving rise to a larger dispersion along 
the flow direction
(see for instance \cite{utter} for a study of Taylor dispersion
in granular materials).
In this case 
it is not possible 
to extract the bare diffusion constant. 
We clearly see the effects of Taylor dispersion 
as a non-Gaussian tail in the distribution.
This leads to a superdiffusive process 
as can be seen from the 
 analysis of the fluctuations of $\Delta \theta$  shown in the inset of
Fig. \ref{sup2}.

\begin{figure}
\centerline{
\resizebox{14.cm}{!}{\includegraphics{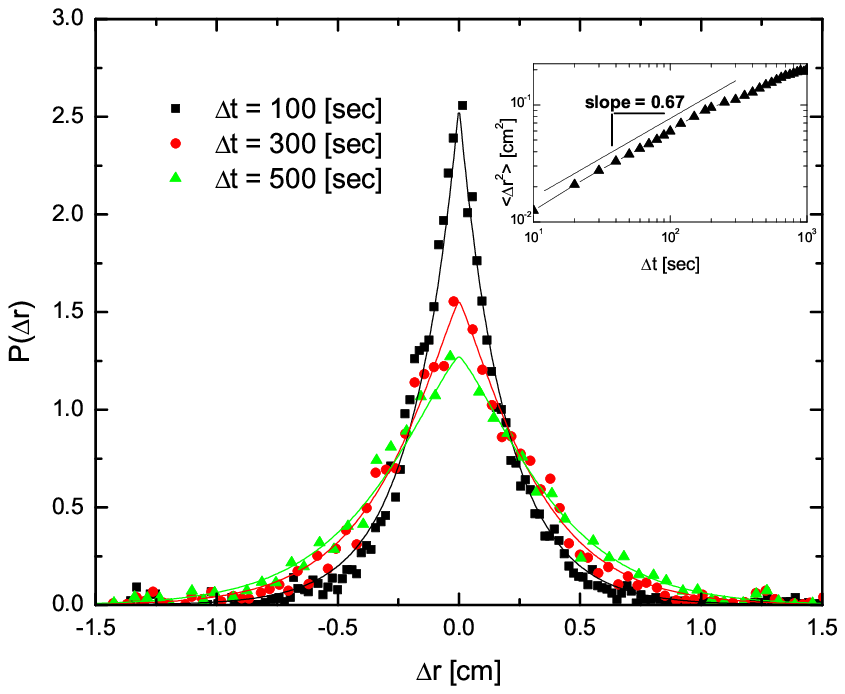}}}
\caption{}
\label{sup3}
\end{figure}

The probability distribution of the 
displacements in the radial direction  $P(\Delta r)$ is shown in 
Fig. \ref{sup3}. 
Interestingly it reveals that the fluctuations 
are stretched exponential with an exponent 
approximately equal to 1.2 (an exponent equal to 2 would correspond
to a Gaussian distribution). 
This result does not invalidate the fact that there could
be a well-defined temperature in the $r$ direction as well.
To calculate such a temperature a force should be applied in the
radial direction, thus its calculation goes beyond the present
experimental set-up.
We also notice  a symmetric
shape for $P(\Delta r)$ indicating the absence of a net flow
of particles towards the inner or the outer cylinder. Therefore
we conclude that 
shear-induced segregation is not present in the
time-scales of the experiment.
An 
average motion of the tracer towards the center of the cell
(a granular Magnus effect) is not observed, either.
The analysis of the fluctuations 
reveals a sub-diffusion process
for short times (of the order of 100 s) shown in the inset.
It should be interesting to apply an external force in the
radial direction to obtain the mobility and then the effective temperature
to test whether the temperature is isotropic
as done in \cite{danna}.

{\bf Summary.---}
We conclude with some remarks: We have tested the existence of the
effective temperature for a given range of particle sizes and
densities.  It remains to be seen how robust our results are under a
more extended set of parameters. These include the use of particles of
different shapes, changing the interstitial liquids, etc.  It would
also be important to test whether the effective temperature of the
packing is the same under different types of driving, for instance
under tapping or shaking, and for different observables.  
We also note that
previous numerical work 
on collective relaxation and self-diffusion in 
granular piles \cite{barker} has showed that in systems 
without Brownian motion the enhanced free volume in the
direction of the perturbation is a more
reliable indicator of diffusion than the tracers particles.
It would be interesting to understand if the analysis of the free volume
gives rise to the same temperature as that measured  here.
Moreover, if
the effective temperature is a proper state variable it should
be independent of the observable. This has been 
previously suggested to be the case in computer simulations 
of effective temperatures of granular materials \cite{ono},
and experimentally it would require to measure
the temperature from, for instance, the 
volume fluctuations.
More experimentation is
needed to fully understand whether the present definition of effective
temperature has a physical thermodynamic meaning.  Our results are a first
step forward in this direction, but do not necessarily imply the validity
of a Boltzmann type of statistics 
for granular materials.


\vspace{1cm}

We are deeply grateful
to M. Shattuck for help in the design of the experiments and
J. Kurchan and  J. Bruji\'c for discussions.
We acknowledge financial support from 
the DOE, Division of Materials
Sciences and Engineering, DE-FE02-03ER46089.

\clearpage

FIG. \ref{couette}. Experimental set-up.
Transparent acrylic grains and black tracers 
in a refractive index and density matched solution
are confined between the inner cylinder of radius $50.8$ mm 
and the outer cylinder of radius $66.7$ mm.

\vspace{1cm}

FIG. \ref{trajectories}.
Trajectories of the 3.97 mm nylon tracers in Packing 1 
showing the diffusion and response to
the gravitational force when sheared in the Couette cell. 
Note that the tracers diffuse distances larger
than the particle diameter indicating that we are probing fluctuations
outside the ``cages'' formed by the surrounding particles.

\vspace{1cm}

FIG. \ref{gauss}. Probability distribution of the displacements of the
3.97 mm nylon tracers in Packing 1. 
The different distributions at different times are
rescaled according to a Gaussian distribution.

\vspace{1cm}

FIG. \ref{mobility}. (a) Diffusion and (b)
mobility of tracers.  We use 
Packing 1 and Packing 2 of 
acrylic particles and tracers of different sizes and densities.
Packing 2 is run at $\dot\gamma_e = 0.024$ 1/s. 
For both packings,  $D$ and $\chi$ are inversely
related to the tracer sizes.

\vspace{1cm}

FIG. \ref{teff}. 
Effective temperatures for various tracers and different
packings as obtained from a parametric plot of their diffusion versus
mobility, as explained in the text. The slopes of the curves for
different tracers consistently yield the same average value of
 $T_{\mbox{\scriptsize eff}} = (1.1 \pm 0.1) \times 10^{-7}$ J
as given by Eq. (\protect\ref{fdt}). 
The inset shows the dependence of 
$T_{\mbox{\scriptsize eff}}$
on the shear rate $\dot\gamma_e$ for the 4.76 mm nylon tracers in Packing 2.


\vspace{.5cm}

FIG. \ref{sup2}. Probability distribution of the angular displacements 
$\Delta \theta$ for
different time intervals. Due to Taylor dispersion effects the distribution 
shows an asymmetric shape. The rms fluctuations shown in the inset reveal
a faster than diffusion process.

\vspace{.5cm}

FIG. \ref{sup3}.
Probability distribution of the displacements 
in the radial direction
$P(\Delta r)$ for different time intervals. A symmetric distribution around
zero displacement indicates 
that there is no net flow in the radial direction. The inset shows the rms
fluctuations.

\clearpage

\clearpage

\clearpage

\begin{figure}
\vspace{1cm}
\centering {\resizebox{14cm}{!}{\includegraphics{makse-fig3.eps}}}
\caption{}
\label{gauss}
\end{figure}

\clearpage

\clearpage

\begin{figure}
\vspace{1cm} \centering {
\resizebox{14cm}{!}{\includegraphics{makse-fig5.eps}}}
\caption{}
 \label{teff}
\end{figure}

\clearpage


\clearpage

\end{document}